\documentclass[twocolumn,secnumarabic,amssymb, amsmath, nofootinbib,tightenlines,
nobibnotes, aps, prl]{revtex4}
\usepackage{amsmath}%
\usepackage{longtable}%
\usepackage{bm}%
\expandafter\ifx\csname package@font\endcsname\relax\else
 \expandafter\expandafter
 \expandafter\usepackage
 \expandafter\expandafter
 \expandafter{\csname package@font\endcsname}%
\fi

\usepackage{graphicx}
\usepackage{graphicx}

\begin{document}

\title{ Anomalous scaling of structure functions and sub-grid models for large eddy simulations of strong turbulence.}

\author{Victor Yakhot$^{1}$ and John Wanderer$^{2}$ \\
$^{1}$Department of Mechanical Engineering\\ Boston
University, Boston 02215\\
$^{2}$EXA Corporation, 55 Network Drive, Burlington, MA 01803\\
}

\begin{abstract} 
 \noindent   The original  goal of Large Eddy Simulations of fully developed turbulent flows  was to accurately describe large-scale flow features  ${\bf u}(\Delta)$ at the scales $r\geq \Delta$ where $\Delta$ is a size of computational mesh. The effect of  small-scale velocity fluctuations ($r<\Delta$)  was  to be accounted for by  effective transport coefficients (subgrid models)  in the coarse-grained Navier-Stokes equations.   It is shown in this paper that,  due to anomalous inertial range  scaling (intermittency) of the moments of velocity difference,  the existing subgrid  models are intrinsically incapable of quantitatively  describing  flow features at  the scales  $r<N\Delta$  with $N\approx 10$.    This increases  computational work approximately  by a factor $10^{3}-10^{4}$. 
The breakdown of the widely used Smagorinsky  relation  for the subgrid  viscosity  on the scales $\Delta/L<1$ is demonstrated and a modification  accounting for  intermittency  of the filtered out small-scale fluctuations is  proposed. 
 \end{abstract}
  
  \maketitle
   
\noindent  {\it Introduction.}   In 1969 S. Orszag introduced spectral methods as a tool for direct numerical simulations  (DNS) of strong turbulence  [1] which were implemented in simulations of three - dimensional isotropic flow in 1970-1971  and published in his paper with Patterson  in 1972 [2]. Thus, a  new field of numerical experimentation    on turbulent flows was born.  Based on Kolmogorov's theory Orszag estimated the scaling of computational work  $W$ with the large-scale Reynolds number  $Re=u_{rms}L/\nu_{0}$: $W=O(Re^{3})$.  Here $L$ and $\nu_{0}$ stand for the integral scale corresponding to the top of inertial  range and molecular viscosity, respectively.  Although it became clear recently  [3]  that, due to intermittency,  the scaling of computational work required for full simulations of developed turbulence can be as bad as  $W=O(Re^{4})$,  this work,  defining computation as an alternative experimental tool for  turbulence studies,  was a major breakthrough.  

\noindent It became clear immediately that due to the unfavorable  Reynolds number scaling of computational work, the  DNS of  high Reynolds number flows of engineering interest were  impractical  and derivation of coarse-grained equations  (models) for prediction of the large-scale flow features, theoretically accounting for the small-scale fluctuations  were  of crucial importance.   The aim  was to eliminate the small-scale velocity fluctuations,   describing structures on the scales $r<\Delta$,  from the Navier-Stokes equations, express their effect in terms of  the resolved field and  reduce the computational work to $W=O(Re^{\gamma})$ with $\gamma\approx  0-1$.    \\

\noindent The problem is formulated as follows.  The Reynolds stress  can be decomposed as:

$$
RS={\bf u\cdot \nabla u}+{\bf u} \cdot \nabla {\bf  u}_{SG}+{\bf u}_{SG}\cdot \nabla {\bf u}+{\bf u}_{SG}\cdot \nabla {\bf u}_{SG}
$$

\noindent where ${\bf u}$ and ${\bf u}_{SG}$ are the resolved and subgrid velocity fields describing flow structures on the scales $r\geq \Delta$ and $r\leq \Delta$, respectively. To fully simulate  velocity fluctuations  on the length-scale $r$  varying  in the interval $L\geq r\geq \eta_{K}\approx LRe^{-\frac{3}{4}}$,  one needs a computational mesh at least as small as  $\Delta\approx  \eta_{K} $.   If  due to computational limitations the practically available mesh size $\Delta\gg \eta_{K}$,  one needs  equations  for velocity field ${\bf u}(\Delta)\equiv {\bf u}$ defined on the resolved scales $r\geq \Delta$. The rapidly varying terms involving the  computationally unaccessible  small-scale subgrid fluctuations  ${\bf u}_{SG}$ on the scales $\eta_{K}<r<\Delta$  are to be averaged out  and their effect represented in terms of the resolved field  ${\bf u}$. \\

First attempts to achieve this goal were made by  Deardorff [4]   who applied the model previously  derived by Lilly and Smagorinsky  [5] to simulations of wall flows.  {\it This model is based on an {\bf assumption} that the effect of small  - scale velocity fluctuations on the large-scale ones can be represented in terms of effective viscosity. It will become clear below that while this assumption is accurate when $\Delta/L\approx 1$,  due  to the anomalous scaling of the inertial range structure functions, it is grossly incorrect when the cut-off $\Delta$ is small, i.e. $\Delta/L\ll 1$. }
The essence of the Lilly-Smagoronsky  construction  can be formulated as follows. Consider a   length - scale $\Delta$  in the inertial range. Then, according to Kolmogorov's  theory, the scale-independent energy flux  

\begin{eqnarray}
\overline{{\cal E}}=\overline{{\cal E}}(\Delta) \approx 2\nu(\Delta)\overline{S^{2}}(\Delta)\approx \nonumber\\
\overline{|(u(x+\Delta)-u(x))^{3}|}/\Delta=\overline{S^{3}(\Delta)}\Delta^{2}
\end{eqnarray}

\noindent where $S(\Delta)\equiv \sqrt{S_{ij}(\Delta)S_{ij}(\Delta)}$,  with $S_{ij}(\Delta)=\frac{1}{2}(\partial_{i}u_{j}+\partial_{j}u_{i})$,  is the rate of strain evaluated on the resolved scales $r\geq \Delta$ and the effective viscosity  $\nu(\Delta)$ accounts  for the contribution  of the small-scale  velocity fluctuations  acting on the scales $r\leq \Delta$.   The filtering of the small-scale velocity fluctuations from the interval of scales $r<\Delta$  generates the resolved field which is smooth on the scales $r\geq \Delta$  and, as a result,  the spatial derivatives  $\partial_{j}u_{i}(\Delta)\approx (u_{i}({\bf x}+\Delta_{j}{\bf j})-u({\bf x}))/\Delta_{j}$  are expressed in terms of velocity increments of the scale $\Delta$.  From here, we have for the well-known relation for turbulent viscosity 
$\nu(\Delta)\approx \Delta^{2} \overline{|S(\Delta)^{3}|}/\overline{S^{2}(\Delta)}$. 

\noindent  This {\bf global}  argument, dealing with the mean properties of a  flow,  is relatively safe. To develop a  numerical method,   one needs  local differential equations for the numerically resolved velocity field with  a {\bf local} expression for turbulent viscosity in terms of  this {\bf resolved}  field.  A  dimensionally correct possibility is the following : based on (1),  write ${\cal E}(\Delta)=2\nu(\Delta)S_{ij}(\Delta)S_{ij}(\Delta)+\Psi(\Delta)$ where $\Psi$ is an unknown function  satisfying  a global constraint $\overline{\Psi}=0$.  {\it Evaluation of this function  which is an extremely difficult task will be discussed below. }   The simplest  and crudest way to develop a model  is by  dropping  the averaging symbols in (1), setting $\Psi=0$  and,  interested in isotropic turbulence, taking   for simplicity  $\Delta_{i}\equiv \Delta$ obtain:

\begin{equation}
\nu_{S}(\Delta)=aS\Delta^{2}\approx a|u(x+\Delta)-u(x)|\Delta
\end{equation}

\noindent which is basically a mixing length model well-known from Prandtl's works. {\it The derivation of this model directly from the Navier-Stokes equation, based on the one-loop renormalized perturbation expansion led to the expression for effective viscosity $\nu_{eff}=\nu_{S}(1-e^{-\frac{\Delta}{\eta_{K}}})$ where $\eta_{K}\approx L Re^{-\frac{3}{4}}$ is the Kolmogorov dissipation scale [6].Typically, in the LES  the cut-off  $\Delta>>\eta_{K}$, so that $\nu_{eff}\approx \nu_{S}$.}
This expression for effective viscosity, combined with the Navier-Stokes equations for the {\bf resolved} velocity field,  is a celebrated Smagorynsky model for the large-eddy-simulations of turbulence (LES).   The expression (2) is the basis  of majority of  modern LES schemes  postulating  the length-scale $\Delta$ identical to the size of a cell of computational mesh.  We can see that since $\nu(\Delta)>>\nu_{0}$, the role of this increased viscosity is to reduce the Reynolds number and, consequently,  the number of  degrees of freedom  necessary for description of the large-scale flow features. In general fluctuating parameter $a$, which is constant in isotropic and homogeneous turbulence cannot be derived from this qualitative argument. 
 
\noindent As follows from (2) the moments of  fluctuating resolved dissipation rate are:

\begin{equation}
\overline{{\cal E}^{n}(\Delta)}=\overline{(\nu_{S}S^{2})^{n}}\propto a^{n}\overline{|(\delta_{\Delta}u|)^{3n}})/\Delta^{n}\propto ({\cal E}L)^{\frac{n}{3}}(\frac{\Delta}{L})^{\xi_{3n}-n}
\end{equation}

\noindent with ``anomalous''  scaling exponents $\xi_{n}$ consistent with the so called Kolmogorov's refined similarity hypothesis.  

\noindent The relation (2) is a result of a general small-scale filtering procedure or, equivalently, the scale elimination methods developed for the dynamic renormalization group.  All existing sub-grid models are based on an {\bf assumption}  that the filtering leads to the equation for the resolved velocity field (for state-of-the-art review,  see Ref.[7])

\begin{equation}
\partial_{t} {\bf u}+u_{i}\nabla_{i}{\bf u}=-\nabla p+\nabla_{\alpha}\nu_{S}(\Delta)\nabla_{\alpha}{\bf u} + EN +{\bf f}
\end{equation}



\noindent    where  $\nu(\Delta)$ originates from the ${\bf u}_{SG}\cdot  \nabla {\bf u}_{SG}$ contributions to the Reynolds stress and 
$EN$ stands for the back scattering term  ("eddy noise")  leading  to a small fraction of energy  the large-scale eddies  receive from interactions  with the small-scale ones.  Interested in three-dimensional turbulence dominated by the direct cascade we,  for a time being,  can neglect the $EN$- contribution.  The large-scale forcing function ${\bf f}$ in the right side of (4) is not modified by the coarse-graining (filtering) procedure described below.  Multiplying (4) by ${\bf u}$,  the energy balance  for the resolved velocity field is calculated readily:

\begin{eqnarray}
{\cal P}=\overline{\bf u\cdot f}=a\Delta^{2}\overline{|S|(\frac{\partial u_{i}}{\partial x_{j}})^{2}}=\nonumber \\
O\big(\frac{\overline{|u(x+\Delta)-u(x)|(u(x+\Delta)-u(x))^{2}}}{\Delta}\big)\nonumber
\end{eqnarray}

\noindent  which, by choosing a proper magnitude for the constant $a$,  is consistent with the Kolmogorov's 4/5 law.  Since by construction,  the arbitrary cut-off $\Delta$  is chosen in the inertial range,  the equation (4) with the Smagorinsky viscosity (2) is expected to produce approximately accurate results for the resolved energy spectrum and the second moment $S_{2}=\overline{(u(x+\Delta)-u(x))^{2}}$.

\noindent However, if consistent, the {\bf local} equation  (4) with sub-grid model (2)  is supposed to describe the resolved velocity field, including all structure  functions $S_{n}(r)=\overline{(u(x+r)-u(x))^{n}}$  and moments of resolved dissipation rate on all scales $r\geq \Delta$.

\noindent  {\it Scale by scale comparison of  LES with DNS }. To assess  the scale-by-scale performance of a  subgrid   model  we have introduced the displacement $r=N\Delta$ where $N>1$ are  the integers,   and evaluated  the modified structure functions  $S_{p}(N\Delta)=\overline{|u(x+N\Delta)-u(x)|^{p}}^{\frac{1}{p}}\equiv \overline{|\delta_{N\Delta}u|^{p}}^{\frac{1}{p}}$ using the DNS (resolution $256^{3}$ ) and  variational LES (resolution $32^{3}$) described in detail in the  Ref.  [8]. For $p=2$ this parameter gives $\sqrt{S_{2}}=(\delta_{N\Delta}u)_{rms}$ and  for an arbitrary $p$ it is a useful tool  for analyzing quality  of prediction  of the most intense small-scale events.  The energy spectra $E(k)$,  evaluated using both methods in Ref.[8],  were in a good agreement with each other [8].  On Fig.1 we show the ratio $S_{p,LES}/S_{p,DNS}$ as a function of displacement $r=N\Delta$.  As expected,  even for this low Reynolds number flow  for which $\Delta/L\approx 0.1$,   the "small-scale"  results rapidly deteriorate with increase of power $p$. This drawback of all existing subgrid models (for different examples see an excellent review [7]) , {\bf which cannot be cured by  introducing the scale-dependent algebraic relations $a=a(r/\Delta)$},  is quite costly:  the model  which is capable of  accurately describing  the flow  on the scales from the interval $r/\Delta\geq 10$ only,  is  at least by the  factor $1000$ more expensive than the one valid in the entire interval $L\geq r\geq \Delta$.     
 

\begin{figure}
\begin{center}
{\includegraphics[width=0.45\textwidth]{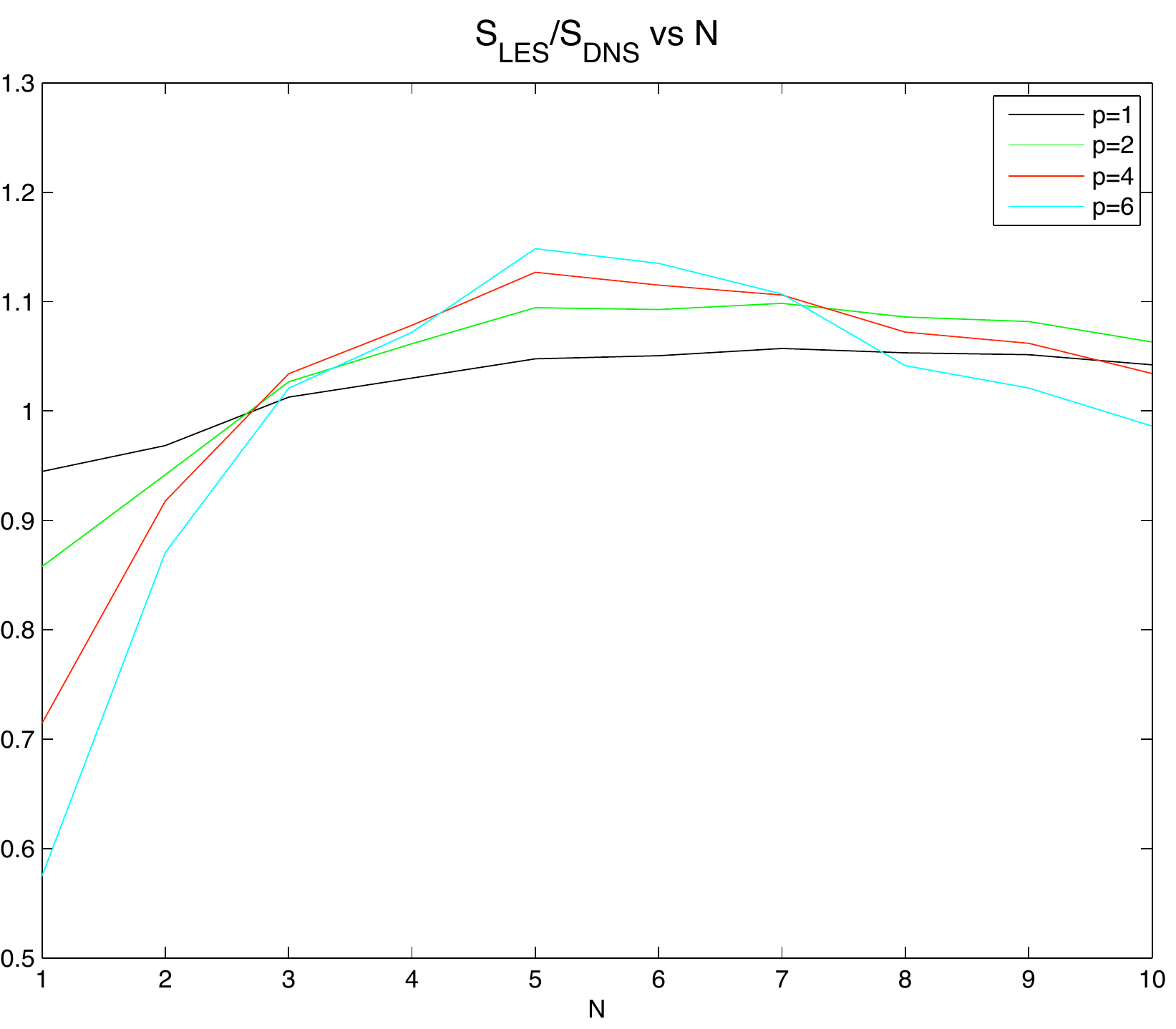}}
\caption{\it    Testing subgrid model:  \   $\frac{S_{LES}}{S_{DNS}}=\frac{\overline{|\delta_{N\Delta}u |^{p}}^{\frac{1}{p}}_{LES}}{\overline{|\delta_{N\Delta}u |^{p}}|^{\frac{1}{p}}_{DNS}}$}
\label{fig1}
\end{center}
\end{figure} 

\noindent  The Smagorinsky model has been derived in Ref. [6] directly from the Navier-Stokes equations from the one-loop renormalized perturbation expansion in powers of the Reynolds number  $Re_{eff}=u_{rms}L/\nu(L)=O(1)$ which can be reformulated as a series in powers of dimensionless rate -of-strain proportional to the velocity increments with displacement $r$ in the inertial range. 
As well-known from experimental data,  at the large scales $r\approx L$,  the velocity fluctuations obey Gaussian statistics and, therefore,  all contributions to the expansion  are of the same order and  a few first contributions to the series give  quantitatively accurate  results.  As $r/L<<1$ this is not so high-order-contributions to the series become dominant and must be taken into account. 
  The severity of this problem, stressed in [9],  becomes clear from a general expression for the probability density of  velocity fluctuations (increments) [10]:

\begin{equation}
P(\delta_{r}u)=\frac{1}{\sqrt{\pi} \delta_{r}u}\int_{-\infty}^{\infty}e^{-x^{2}}dx\int_{-\infty}^{\infty}e^{in\ln\frac{ r^{a}\sqrt{2}x}{\delta_{r}u}-n^{2}\ln r}dn
\end{equation}

\noindent where, to simplify notation,  we set $r\equiv r/L$ and $ \delta_{r}u\equiv \delta_{r}u/u_{rms}$. In the limit $r\rightarrow 1$, this expression gives Gaussian PDF $P\propto e^{-(\delta_{r}u)^{2}/2}=e^{-\frac{u_{sp}}{u_{rms}}^{2}}$  which is a probability density of single-point  velocity field.

\noindent  The small-scale  $(r/L<1)$  strong luctuations  are well described  by the expression following from the steepest descent evaluation of the integral (5) (for details see [10]-[11]):

\begin{equation}
P(\delta_{r}u)=\frac{2}{\pi (\delta_{r}u)\sqrt{4|\ln r^{b}|}}\int_{-\infty}^{\infty}e^{-x^{2}}Exp[-\frac{(\ln \frac{\delta_{r} u}{r^{a}\sqrt{2x}})^{2}}{4b|\ln r|}]dx
\end{equation}

\noindent  The constants in (5)-(6)   tested in various flows [11]-[13],  are $a\approx 0.383$ and $b\approx 0.01667$.   As   $\Delta/L\rightarrow 0$, the PDF  (6) is an extremely shallow function stressing the importance of strong small-scale fluctuations not accounted for by  Smagorinsky-like  models.   Thus, to eliminate small-scale fluctuations $r<\Delta$, one has to use (5) -(6) as a filter  applied to the Navier-Stokes equations which is not a simple task. \\

\noindent 
To illustrate the role of intermittency (anomalous scaling) in the LES modeling we, can use an  approach leading to  derivation of turbulence models  from a  kinetic  equation  for some artificial  ``particles''  forming a ``fluid'' characterized  by  the relaxation time $\tau$   reflecting dynamic properties of a  system.  For example, choosing  $\tau=const$ leads  in the first order of the Chapman-Enskog expansion
to the Navier-Stokes equations  with viscosity $\nu_{0}=\tau u_{rms}^{2}/3$  with non-linear corrections appearing from  the higher order terms.  The same procedure but with the relaxation time  $\tau\propto {\cal K}/{\cal E}$  yields  the well-known linear and non-linear ${\cal K}-{\cal E}$ models for the large-scale velocity field [14].  Repeating the procedure,  presented in Refs.[14],  it is easy to see that the LES, Smagorinsky-like,  models are generated by the same equations   with  $\tau(\Delta)=1/\overline{|S_{ij}(\Delta)|}$. 

\noindent  In the second order of perturbation expansion in powers of dimensional rate of strain the procedure gives:

\begin{eqnarray}
\sigma_{ij}=2\nu_{S}(\Delta)S_{ij} -2\nu_{S}(\Delta)(\partial_{t}+
{\bf u\cdot\nabla})(\tau S_{ij})/3\nonumber\\
-4\nu_{S}(\Delta)\tau \big[ S_{ik}S_{kj}-
\frac{1}{d}\delta_{ij}S_{kl}S_{kl}\big]+\nonumber\\
\nu_{S}(\Delta)\tau[S_{ik}\Omega_{kj}+S_{jk}\Omega_{kj}]
\end{eqnarray}

\noindent with  $\nu_{S}=\tau u^{2}_{rms}(\Delta)=\Delta \frac{\overline{u(\Delta)^{2}}}{\overline{|\delta_{\Delta}u|}}=a\Delta^{2}|S|$.   The same relation has been derived by Rubinstein et al [15] directly from the Navier-Stokes equations using scale-elimination technique and,  even earlier,  by Speziale [16] who applied symmetries used in rheology for derivation of turbulence models.   {\it On the first glance  dropping the averaging symbol from expression $\tau u^{2}_{rms}(\Delta)$, which is used for derivation in the  procedure leading to Smagorinsky model,  cannot be accurate. However, a consistent derivation leading to the  expression (7) involves averaging over relatively fast velocity fluctuations varying on the time scales  $\tau(r<\Delta)<\tau(\Delta)$. Thus the velocities on the scale $r\approx \Delta$ can approximately be considered  "slow modes". This  observation somewhat justifies an assumption leading to (7).} As we see, the Smagorinsky model, which is the basis of all modern numerical schemes for LES is  obtained in the lowest order of expansion in powers of dimensionless rate of strain. Now, we will discuss the limits of its validity.

 Our goal is to express the sub-grid contribution to the Reynolds stress is terms of the resolved field ${\bf u}(r>\Delta)$.  As one can see from (7), due to prolifiration of the subscripts,  with increase of the power of expansion  each term becomes more and more  comlex.  Some simplifications are based on the following: The resulting equation must be invariant under Galileo transformation.  Only two dimensionless operators   satisfy this requirement: $D_{1}=\tau(\partial_{t}+u_{i}\nabla_{i})$, responsible for the memory effects,  and $D_{2}=\tau\frac{\partial u_{i}}{\partial u_{j}}$.  Thus, the general expression for the Reynolds stress is
$\sigma_{ij}=\sigma_{ij}(D_{1},D_{2})$.

\noindent It has been shown [17]-[18] that expansion in powers  of $D_{1}$ results in the  telegrapher equation with O( $\tau D_{1}^{2}$)-term  in  the left side of  (4, )  responsible for the quasielastic response of turbulent flow to external high-frequency perturbations.  In what follows, mainly interested in the intermittency contributions to $\sigma_{ij}$, we omit this term which is important for the flows rapidly varying  on the time scale $\tau<<\tau(\Delta)$.
 To simplify the treatment and to avoid dealing with  an extremely complex tensorial structure of  high-order terms in the expansion (7)  we,  based on the constraints discussed above,   choose  an infinite  subset of  leading contributions to  each term in power  series  of $\sigma_{ij}$   similar  to the "single-back-bone"  diagrams of the Wyld's  expansion of the Navier-Stokes equations:

\begin{equation}
\sigma_{ij}\approx \nu_{S} S_{ij}(\Delta)\sum_{n}\alpha_{n}\Big(\frac{|S_{ij}(\Delta)|}{\overline{|S_{ij}(\Delta)|}}\Big)^{n}
\end{equation}

\noindent  where the coefficients $\alpha_{n}$ can be evaluated only for the first few terms.  {\it Similar approach has been developed for the scalar problem in Ref.[19],  where it was used for assessment of the accuracy of the $\epsilon$-expansion} in turbulence theory.  It is clear that the mean dissipation rate  $\overline{{\cal E}}=
-\overline{u_{i}\frac{\partial \sigma_{ij}}{\partial x_{j}}}$
and the sub-grid viscosity  can be defined as:

\begin{equation}
\nu_{SG}(\Delta)=\sigma_{ij}S_{ij}^{-1}=\nu_{S}\sum_{n}\alpha_{n}\Big(\frac{|S_{ij}(\Delta)|}{\overline{|S_{ij}(\Delta)|}}\Big)^{n}
\end{equation}

\noindent  This expression is easily understood  if we notice that the  expansion we are dealing with is in powers of dimensionless rate-of-strain $\eta=\tau(\Delta)S$ similar to that discovered in derivation of the RNG ${\cal K}-{\cal E}$ model from the Navier-Stokes equations [20]. Adding to this expression  contributions including $\Omega_{ij}$ does not change  qualitative conclusions  of the argument.

\noindent As we see,  in case of normal scaling or in the large-scale limit  $\Delta/L\rightarrow 1$,  the ratios 
$\overline{\Big(\frac{|S_{ij}(\Delta)|}{\overline{|S_{ij}(\Delta)|}}\Big)^{n}}=\beta_{n}$  are $\Delta$-independent and  the dissipation rate based on this sub-grid viscosity  in the limit $\Delta\rightarrow L$ is proportional to 

\begin{eqnarray}
\overline{{\cal E}}(L)\propto  \overline{\nu_{SG}(L)S^{2}(L)}
\approx \frac{\overline{|\delta_{L}u|(\delta_{L}u)^{2}}}{L}\sum_{n}\alpha_{n}\beta_{n}\approx \overline{\cal E}={\cal P}
\end{eqnarray}

\noindent  provided  $\sum_{n}\alpha_{n}\beta_{n}$ is a finite number.   This expression is equivalent to Smagorinsky model (2).     In the large-scale limit $\Delta\rightarrow L\rightarrow\infty$, due to the Galileo invariance,  the corrections to the non-linear term  ${\bf u\cdot \nabla u}$  in the coarse-grained equation (4)  are equal to zero  and,    as $\Delta/L\approx 1$,  the energy balance in the equation (4) with viscosity given by (9) is not violated.

A completely different result is obtained  when anomalous scaling is taken into consideration on the scales $\Delta<L$.  It is clear from (9) that  in  this case the dissipation is not equal to (10) but  is multiplied by  an infinite sum:
\begin{equation}
\overline{{\cal E}}(\Delta)\approx \overline{\nu_{SG}(\Delta)S^{2}(\Delta)}\approx \overline{{\cal E}}(L)\sum_{n}\alpha_{n}\beta_{n} (\frac{\Delta}{L})^{\xi_{n+3}-\frac{n}{3}-1}\approx {\cal P}
\end{equation} 

\noindent Since $\xi_{n+3}-\frac{n}{3}-1<0$,  for  $\Delta/L<<1$ the dominating contributions from the high-orde terms in (10)  invalidate the Smagorinsky's model  in the range of scales  $\Delta/L<<1$.   As we see, since production ${\cal P}$ is a large-scale  $\Delta$-independent property,  when  $\Delta/L<<1$,  the energy balance in the LES equation (4) can be restored only by renormalization of the non-linear term, i.e. by accounting for the high-order non-linear contributions,  disappearing in the limit $\Delta/L\rightarrow 1$.   In this case the relation proportional to (9) is a possible candidate satisfying the above requirements.  Thus,  neglecting  the ``rapid-distortion'' (memory) contribution,  the  LES models accounting for the inertial range intermittency is:

\begin{equation}
\partial_{t} {\bf u}+\Phi\{S,\frac{\Delta}{L}\}u_{i}\nabla_{i}{\bf u}=-\nabla p+\nabla_{\alpha}\nu_{SG}(\Delta)\nabla_{\alpha}{\bf u} + EN +{\bf f}
\end{equation}

\noindent  As $\Delta/L\rightarrow 1$ the functional $\Phi\rightarrow 1$.  In the interval $\Delta<L$ it ensures  cancellations leading to the cut-off -independent expression for the mean dissipation rate  $\overline{{\cal E}}(\Delta)=\overline{{\cal E}}={\cal P}$.   

\noindent To conduct numerical simulations  using equations with transport coefficients given  by  infinite series in powers of  dimensionless rate-of-strain $\tau S$ is not a simple task. One can attempt to qualitatively represent the sum in (9) in a compact form like $\nu_{SG}(\Delta)=\nu_{S}/(1+\frac{S^{2}}{\overline{S}^{2}_{1}})$ which,  by  the  averaging the expression for the dissipation rate using the probability density (5)-(6) is equivalent to the infinite series (11).   This qualitative representation of an infinite series (9) proved to be very useful in the ${\cal K}-{\cal E}$ modeling emerging from a present theory in the limit $\tau(\Delta)\rightarrow \tau(L)\propto {\cal K}/{\cal E}$ [21]. 

\noindent To summarize: in this paper  we compare  the LES based on Smagorinsky sub-grid model with the direct simulation of the same flow.  It  has  been demonstrated that,  while the LES gives  a reasonably accurate representation of the large-scale flow features, it completely underestimates strong small-scale  ($r<10<\Delta$) intermittent velocity fluctuations. The reason for this failure is traced to the inadequate account for the small-scale intermittemcy of  the filtered out small-scale fluctuations in the Smagorinsky-class models. An improved model including the intermittent phenomena is proposed.

\noindent It is interesting to compare  the outcomes of this work with the results presented in two important papers [22]-[23].   In these works the authors compared filtered and unfiltered experimental data on turbulent properties of a few flows (wakes, grid, decaying turbulence, etc).  The filtering procedure of Refs.[22]-[23] was the same as the one used in the LES.  Since no modeling was involved,  and one would expect that the resolved velocity fluctuations on the scales $r\leq \Delta$ of the two fields  must be identical.  A completely different picture emerged:  the strong small-scale features of the filtered field were grossly  underpredicted  compared   with  the raw, unfiltered,  data.  This important result  indicates  that large and small -scale velocity fluctuations strongly interact and one cannot obtain a reasonable representation of one by simply filtering out the other. 
This conclusion  becomes clear if one recalls  that  in turbulent flows extremely  thin and long geometrical structures,  like pancakes, ``worms'' etc,  consist of  inseparable,  strongly interacting  large and small-scale flow features.  Thus,  smoothing the velocity field over length-scale $\Delta$ (filtering), distorts the entire field including the interval $\Delta<r<10\Delta$.  In short, the  works [22]-[23] showed that the simple-minded  filtering procedure is intrinsically flawed.  The theory   described in this paper, based on consistent procedure of small-scale elimination from the Navier-Stokes  or Boltzmann-BGK equations,  proposes a remedy  for this drawback  by strong modification of the equations  of motion for the resolved scales. The resulting model, while at the large scale equivalent to the ordinary Smagorinsky model,  at the small scales $r\approx \Delta$ does not resemble neither it nor the Navier-Stokes equations themselves .

\noindent Correctly accounting for the entire range of velocity fluctuations $L>\Delta>\eta_{K}$  may substantially  reduce cost of numerical simulations  and  improve  quality of LES of mixing and sound generation by turbulence. This will be a subject of  future communications.\\

 \noindent {\bf References.}\\
1.\  S.A. Orszag, ``Numerical methods for the simulation of turbulence'', Phys.Fluids {\bf 12} 250 (1969).\\ 
2. \ S.A. Orszag and  G.S. Patterson, ``Numerical Simulation of Three-Dimensional Homogeneous Isotropic Turbulwence'', Phys.Rev.Lett.{\bf 28}, 76 (1972).\\
3. \ V. Yakhot,  ``Pressure-velocity correlations and scaling exponents in turbulence'', J.Fluid Mech. {\bf 495} (2003); \\
4. \ J.W. Deardorff, J. Fluid Mech. {\bf 41},453 (1970); J.W.  Deardorff, J.Comp.Phys.{\bf 7}, 120 (1971); \\
5. \  J. Smagorinsky,  Monthly Weather Rev.  {\bf 91},  99 (1963);\\
6. \  V.Yakhot  and S.A. Orszag, ``Renormalization group analysis of turbulence", J. Sci. Comp. {\bf 1},  3, (1986).\\
7. \  P. Sagaut, ``Large Eddy Simulation of Incompressible Flows: An Introduction'', Springer (2005) \\
8. \ J. Wanderer  and A.A. Oberai, ``A Two-Parameter Variational Multiscale Method for Large Eddy Simulations'',  Phys.Fluids.{\bf 20}, 085107 (2008)\\
9 . \  V. Yakhot and K.R. Sreenivasan, "Anomalous scaling of structure functions and turbulence simulations", J. Stat. Phys. {\bf 121}, 823 (2005);\\
10. \ V. Yakhot, ``Probability Densities in Strong Turbulence'', Physyca D{\bf 215}, 166 (2006);\\
11. \  J. Schumacher, K.R. Sreenivasan and V. Yakhot,  ``Asymptotic exponents from low-Reynolds number flows''\, New  J. Physics {\bf 9}, 89 (2007).\\
12. \  S.C.C. Bailey, M. Hultmark, J. Schumacher, V. Yakhot, A.J. Smits, ``Measurements of local dissipation scales in turbulent pipe flow'', Phys.Rev.Lett. {\bf 103}, 14502 (2009);\\
13. \  J. Schumacher, private communication 2011;\\
14. \  \ H. Chen, S. Orszag, I. Staroselsky \& S.Succi, ``Extended analogy between Boltzmann  kinetic theory of fluids and turbulence", J.Fluid Mech.  {\bf 519}, 301 (2004).\\
15. \  R. Rubinstein  and J.M. Barton, `` Renormalization group analysis of the Reynolds stress transport equation'', Phys.Fluids A{\bf 4}, 1759 (1992). \\
16. \  C. G. Speziale, ``On non-linear $K-l$ and $K-\epsilon$ models of turbulence", J.Fluid.Mech {\bf 178}, 459 (1987).\\
17.  \  V. Yakhot \& C. Colosqui, ``Stokes' second problem in the high-frequency limit: application to nanomechanical resonators", J.Fluid.Mech. {\bf 586}, 249 (2007).\\ 
18. \  H. Chen, S.A. Orsazg and I. Staroselsky, ``Macroscopic description of arbitrary Knudsen number flow using Boltzmann-BGK kinetic  theory'', J.Fluid Mech. {\bf 658}, 294, (2010). \\ 
19. \  S.A. Orszag and V. Yakhot, ``Analysis of ${\epsilon}$-expansion in turbulence theory'', J.Sci.Comp. {\bf 14}, 147 (1999);\\
20. \  V. Yakhot, S.A. Orszag, T. Gatsky, S. Thangam \& C.Speciale, ``Development of turbulence models for shear flows using a double expansion technique", Phys. Fluids A{\bf 4}, 1510  (1992).\\ 
21. \   H. Chen et.al , ``Extended Boltzmann Kinetic Equation for Turbulent Flows'', Science {\bf 301}, 633 (2003);\\
22. \  S. Cerutti and C. Meneveau, `` Statistics of filtered velocity in grid and wake turbulence'', Phys. Fluids {\bf 12}, 1143 (2000);\\
23. \ H.S. Kang, S. Chester and C. Meneveau, ~~Decaying turbulence in an active -grid-generated flow and comparisons with large-eddy-simulations'', J.Fluid Mech {\bf 480}, 1145 (2000);\\

 \end{document}